\documentclass[10pt,a4paper]{article}

\usepackage{amsmath,amsthm,amssymb}
\usepackage{graphics,graphicx}
\usepackage{epsfig}
\usepackage{multicol}
\usepackage{color}
\makeatletter
\@addtoreset{equation}{section}
\makeatother

\begin{document}
\begin{flushright}
QMUL-PH-07-11\\
RH-04-2007
\end{flushright}
\begin{center}
\Large\bf{Type I singularities and the Phantom Menace}
\end{center}
\begin{center}
\textbf{Tapan Naskar} ${}^{a,}$\footnote{tapan@iucaa.ernet.in} \textbf{and John Ward} ${}^{b,c,}$\footnote{j.ward@qmul.ac.uk} \\
\end{center}
\begin{center}
\it{${}^a $IUCAA, Post bag 4,
Ganeshkhind, Pune 411007, India} \\
\vspace{0.2cm}
\it{${}^b$ Center For Research in String theory, Department of Physics, \\
Queen Mary
University of London, Mile End Road, London, E1 4NS, UK.}\\
\vspace{0.2cm}
\it{${}^c $ Science Institute, University of Iceland, Taeknigardi, Dunhaga 5 IS-107 Reykjavik, Iceland.}
\end{center}
\begin{abstract}
We consider the future dynamics of a transient phantom dominated phase of the universe in LQC and in the RS braneworld, which 
both have a non-standard Friedmann equation. We find that for a certain class of potentials, the Hubble parameter oscillates with simple harmonic motion in the LQC case and therefore avoids any future singularity. For more general potentials we find that damping effects
eventually lead to the Hubble parameter becoming constant. On the other hand in the braneworld case
we find that although the type I singularity can be avoided, the scale factor still diverges at late times.
\end{abstract}

\section{Introduction}
Observations suggest that the universe is currently dominated by a dark energy phase, which accounts for the apparent acceleration
\cite{observations}. However this cannot be reconciled within conventional cosmology, and therefore provides an important
theoretical challenge \cite{Sahni:1999gb} . One possible
explanation for this dark energy is that there exists a non-zero cosmological constant, implying that we are currently in a de Sitter
phase. The origins of such a constant have been the focus of much recent work in flux compactifications of superstring and M-theory (see \cite{stringcompactifications} and the references therein.).
An alternative proposal is that the energy density of the universe is currently dominated by a phantom scalar field, which
has a canonical scalar Lagrangian albeit with the wrong sign in the kinetic term \cite{phantoms}. This leads to a violation of the weak energy condition such that $p < \rho$. 
The physical origins of such a field, however, remain problematic at best since the Hamiltonian is unbounded from below which means that a stable vacuum state may not exist upon quantization of the theory. In addition is was shown that phantom fields must have minimal, or zero, interactions with other matter - and that they themselves must only interact gravitationally \cite{cline}. However see \cite{stablephantom} for a potential resolution to the phantom problem. 

One resolution to this problem is that the phantom fields may only exist temporarily, which is what we will assume in this note (although we should point out that these kind of fields can naturally arise within certain string theories, most notablytype II* string theory \cite{stringghosts}). Let us imagine that the kinetic term in the Lagrangian has some moduli space metric $G_{ij}(\phi)$ multiplying the derivative terms such that we have a term $G_{ij} \partial_{\mu} \phi^i \partial^{\mu} \phi^j$. 
Usually one takes the metric to be positive definite, however let us assume that there is some region of moduli
space where it has a negative eigenvalue. Thus any fields $\phi^i$ entering this region will have negative kinetic energy, at 
least for the duration of time that the field find itself in this region of moduli space \cite{sub}. It is \emph{this} region that we shall 
consider in this note which therefore implies that we expect the current dark energy phase to be a transient phenomenon. See \cite{ish, vikman} for related issues. Of course the scalar field may be remain in this region for a long time, and thefore the accelerating phase may appear to be long lived. 

In a typical Friedmann-Robertson-Walker (FRW) model exhibiting a dark energy, or phantom phase, such a universe will contain a future singularity.
In this type of singularity, referred to as a Type I singularity \cite{typeIsingularity}, both the energy density $\rho$ of the dark
energy and the Hubble parameter diverge within a \emph{finite time}. Clearly this is undesirable from a physical perspective, and so
many authors have tried to resolve these future singularities \cite{avoidance}. Although there has been much progress in resolving initial singularities
in string theory, future singularities are problematic. This is because the low energy solutions of perturbative string theory simply 
match onto the standard FRW cosmologies. Therefore we are motivated to move beyond this approach to search for other solutions.
Of course we may find additional corrections once a fully consistent non-perturbative description has been realized.

An alternative approach to quantum gravity is via loop quantum gravity. In this theory non-perturbative effects lead to $\rho^2$ corrections to the standard Friedmann equation and thus allow us the possibility of resolving any future singularities \cite{loops, Ashtekar:2006rx,Singh:2006sg,Ashtekar:2006bp}.
It is this non-canonical dependence upon the energy density that leads to interesting physical behavior, particularly
when the energy density is near the critical density $\rho_c$. As such backgrounds of this type deserve further investigation.
Recently it was observed that these non-perturbative corrections can lead to the future oscillation of $H$ between finite values, implying that
neither the scale factor nor the energy density diverge. Instead both parameters become oscillatory and bounded at finite value \cite{Sami:2006wj}, thus avoiding any future singularities.

Another interesting correction to the Friedman equation can be seen in studies of braneworld cosmology \cite{braneworlds}, where
there is a $\rho^2$ correction which has the opposite sign to the loop background. In transpires that there is an interesting and highly
non-trivial duality between these solutions \cite{Singh:2006sg} suggesting that the two models are related to one another.

In this note we will study the qualitative differences between the loop and braneworld inspired solutions, and their
potential consequences for the evolution of the universe. We will see that in the loop case the Hubble parameter behaves as
a damped harmonic oscillator with frequency $\omega$, however in the brane case the oscillation frequency rapidly becomes complex
and the Hubble parameter is frozen at a constant value Although this avoids the type I singularity, the scale factor continues to diverge
and therefore we do not resolve the singularity issue.

Note: Upon completion of this work we learnt of the paper by \cite{burin} which investigates a similar issue. The conclusions reached there
are similar to those presented here. In both cases we emphasis that the Hubble parameter can become oscillatory, although our analysis shows that this will only be true for a specific class of phantom potentials. We also highlight the similarity between this model, and that
coming from a braneworld motivated one.
\section{Phantom Scalar Fields and Loop Backgrounds.}
The energy density $\rho$ and pressure $p$ of a phantom field are defined by the following components of the energy momentum tensor
\begin{eqnarray}
\rho&=&-\frac{1}{2}\dot{\phi}^2+V\label{rho} \nonumber \\
p&=&-\frac{1}{2}\dot{\phi}^2-V\label{p}
\end{eqnarray}
and the equation of motion
\begin{equation}
\ddot{\phi}+3H\dot{\phi}=V'\label{KG}.
\end{equation}
This is enough to specify the dynamics of the minimally coupled phantom field coupled to standard Einstein gravity.
The standard Friedmann equation, in terms of the reduced Planck mass $M_p^2=8 \pi G$, is simply
\begin{equation}
H^2 = \frac{\rho}{3 M_p^2} \nonumber
\end{equation}
however non-perturbative loop quantum gravity effects modify this to read\footnote{This equation also arises in braneworld constructions with an additional time-like direction, as shown in Shtanov and Sahni \cite{braneworlds}.}
\begin{equation}
H^2=\frac{\rho}{3M_{p}^{2}}\left(1-\frac{\rho}{\rho_c}\right)\label{friedman}
\end{equation}
where $V'\equiv dV/d\phi$ and $\rho_c\equiv \sqrt{3}/(16\pi\gamma^3G^2\hbar)$ is the critical energy density. Here $\gamma$ is the Barbero-Immirizi parameter which arises from the Poisson bracket of the conjugate connection $c$ and the triad $p$, through the relation
$\lbrace c,p \rbrace= 8\pi G \gamma/3$. Numerically $\gamma$ is typically fixed to be small. We refer the interested reader to \cite{loops} for a detailed account of the quantization procedure for such a theory.
Combining these we can write the derivative of $H$ with respect to time
\begin{eqnarray}
\dot{H} &=& -\frac{\biggl(\rho+p\biggr)}{2M_p^2}\Biggl( 1- \frac{2\rho}{\rho_c}\Biggr).
\label{hdot}
\end{eqnarray} 
As usual in these problems it is easier to work in terms of dimensionless variables, so we make the following definitions
\begin{eqnarray}
\tilde{H} &=& \sqrt{\frac{3M_p^2}{\rho_c}}H, \hspace{1cm}
\tilde{t} = \sqrt{\frac{\rho_c}{3M_p^2}}t \nonumber \\
\tilde{\phi} &=& \frac{\phi}{\sqrt{3M_p^2}}, \hspace{1.2cm}
\tilde{V} = \frac{V}{\rho_c} 
\end{eqnarray}
and using these dimensionless quantities we see from (\ref{friedman}, \ref{hdot}) that the dimensionless equations of motion become.
\begin{eqnarray}\label{eq:dimlesseom}
\tilde{H}^2 &=&\Biggl(-\frac12 \biggl(\frac{d\tilde{\phi}}{d\tilde{t}}\biggr)^2 + \tilde{V}\Biggr)
                \Biggl(1+\frac12\biggl(\frac{d\tilde{\phi}}{d\tilde{t}}\biggr)^2 -\tilde{V}\Biggr) 
                \nonumber \\
\frac{d \tilde V}{d\tilde \phi} &=&\frac{d^2\tilde{\phi}}{d\tilde{t}^2} + 3 \tilde{H}\frac{d\tilde{\phi}}{d\tilde{t}} \nonumber \\
\frac{d\tilde{H}}{d\tilde{t}}&=& \frac32\biggl( \frac{d\tilde{\phi}}{d\tilde{t}}\biggr)^2 \Biggl(1+\biggl(\frac{d\tilde{\phi}}{d\tilde{t}}\biggr)^2 -2\tilde{V}\Biggr)
\end{eqnarray}
We want to re-write these expressions in an autonomous form, so it is convenient to define the following variables
\begin{eqnarray}
x &\equiv& \frac{d\tilde{\phi}}{d\tilde{t}}, \hspace{1.8cm}
y \equiv \sqrt{\tilde V} \nonumber \\
\lambda &\equiv& -\frac{1}{\tilde{V}}\frac{d\tilde{V}}{d\tilde{\phi}}, \hspace{1cm}
\Gamma \equiv \frac{VV''}{V'^2}
\label{def}.
\end{eqnarray}
And thus the autonomous equations of motion can be written 
\begin{eqnarray}
\frac{dx}{d\tilde{t}} &=& -3\tilde{H}x - \lambda y^2 \nonumber \\
\frac{dy}{d\tilde{t}} &=& -\frac12\lambda xy \nonumber \\
\frac{d\lambda}{d\tilde{t}} &=& \lambda^2\biggl(1-\Gamma\biggr)x.
\end{eqnarray}
Additionally we see that the Friedmann equation in dimensionless variables becomes simply
\begin{eqnarray}
\tilde{H}^2 &=& \Biggl(-\frac12x^2+y^2\Biggr)\Biggl(1+\frac12x^2-y^2\Biggr).
\end{eqnarray}
The non-negativity of this last expression implies that $(x^2/2-y^2) \in [-1,0]$ or in terms of the variable $x^2$ we see that
this must lie in the closed set $x^2 \in [2(y^2-1), 2y^2]$. The variables are constrained, and not independent which is why the Hubble parameter never becomes imaginary even when $x=0$. This can be
demonstrated graphically using a Monte Carlo simulation and shown in the figure below

\begin{figure}[ht]
\centering
\includegraphics[width=10 cm,angle=-90]{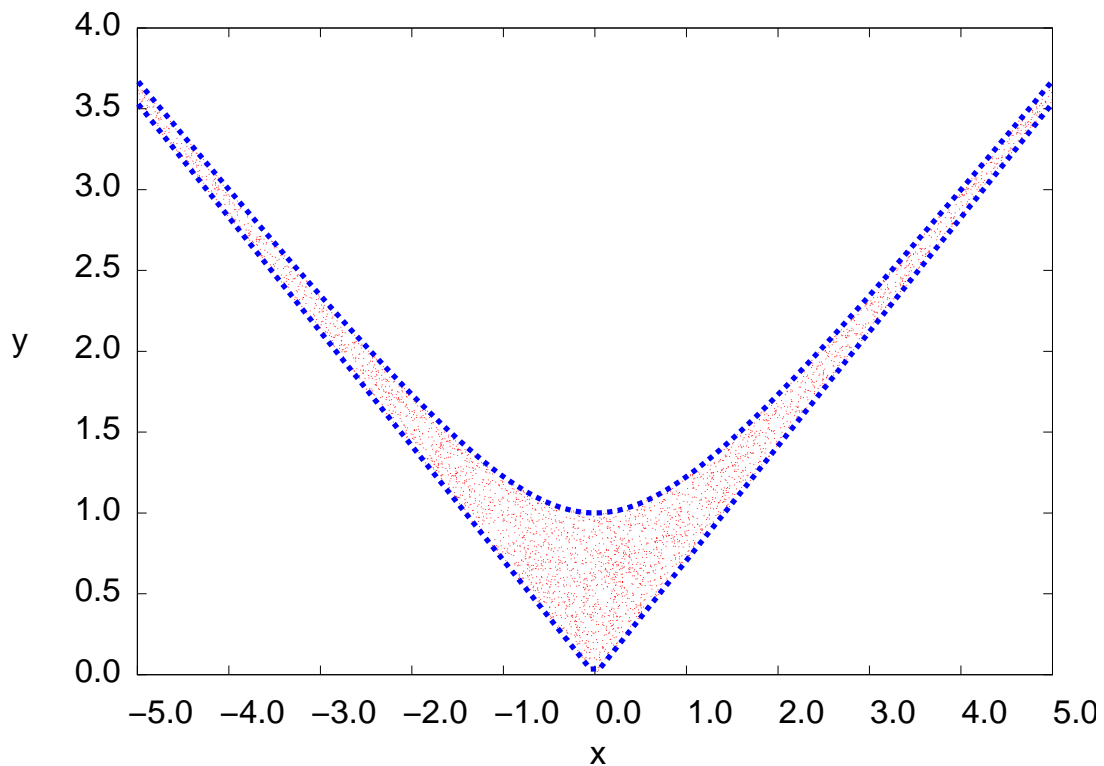}
\caption{Phase space scatter plot of the $\lbrace x,y \rbrace$ plane generated using a Monte Carlo algorithm. The boundary (blue lines) 
defines the region of allowed points.}
\end{figure}

As we move away from the origin we see that $x$ becomes proportional to $y$ as expected from our analytic argument.
Since $y^2$ is the dimensionless potential for the phantom field this indicates that we should find scaling solutions when $y^2>>1$, so that we can neglect all $1/y^2$ terms as being negligibly small.

Importantly we can write the acceleration of the Hubble parameter in the following form using (\ref{eq:dimlesseom})
\begin{eqnarray}
\frac{d^2 \tilde{H}}{d\tilde{t}^2} &=& -9x\Biggl(x+2x^3-2xy^2\Biggr)\tilde{H}-3\lambda y^2 \Biggl(x + x^3 - 2xy^2\Biggr) \nonumber\\
                                   &=& -\omega^2 \tilde{H} - \gamma \frac{d\tilde{H}}{d\tilde{t}} 
\end{eqnarray}
Where we have defined 
\begin{eqnarray}\label{eq:acceleration}
\omega ^2 &=& 9x^2\Biggl(1+2x^2-2y^2\Biggr) \\
\gamma    &=& 2\lambda \frac{y^2}{x} 
\end{eqnarray}
We see that $\tilde{H}$ is a damped, harmonic oscillatory function with frequency $\omega$ and damping factor $\gamma$. This is
an important result, since this implies that $H$ can oscillate and therefore avoid future singularities.
The damping factor scale is set by $\lambda$, and decreases as $y$ increases, therefore one would expect the damping to
die away as a function of time. The frequency is an ever increasing function of $y$ provided that $y^2 > 3/2$, which is a reality
condition. One can see that the frequency will never become complex, because the value of $y$ is constrained by the value of $x$. 
Again we can confirm the validity of said assumption using Monte Carlo techniques

\begin{figure}[ht]
\centering
\includegraphics[width=10 cm, angle=-90]{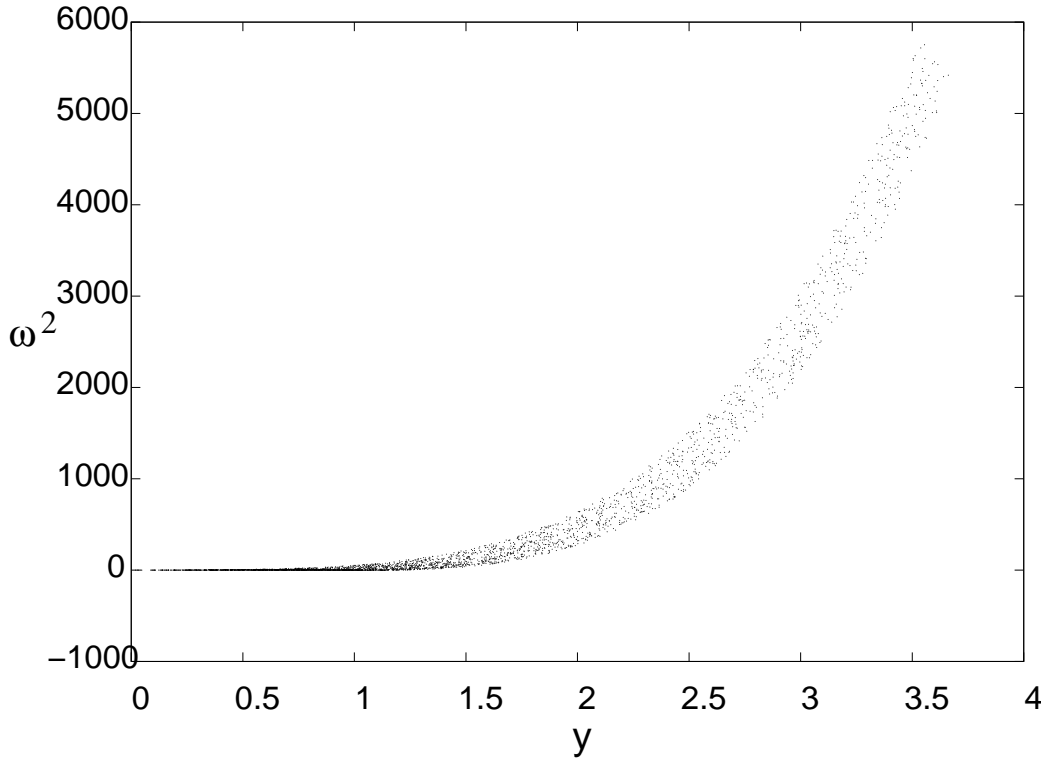}
\caption{Plot of the square of the frequency as a function of $y$, using the causality constraint. The scattered points
represent the allowed values of $\omega^2$. It is noticeable that there is a band of physical solution which increases in
width as $y$ also increases.}
\end{figure}

Therefore it would appear that at late times $H$ oscillates more rapidly, becoming a simple harmonic oscillator.
Schematically we see that the Friedmann equation looks like $H^2 \sim \tilde\rho(1-\tilde\rho)$, and so we deduce that 
$\tilde{\rho}$ must also be an oscillatory function.

We can define an effective equation of state for the phantom field in the usual manner
\begin{equation}
\Omega_{\phi} = \left(-\frac{x^2}{2} -y^2 \right) \left(-\frac{x^2}{2} + y^2 \right)^{-1}
\end{equation}
which essentially tells us that phantom dominated solutions require $x^2 > 0$ since all dependence on the potential vanishes
when we look at violations of the weak energy principle.
Despite this, we know that the kinematics of the phantom field are determined by the form of the scalar potential. Therefore
we should try to analyze the equations of motion with this in mind.
There are essentially three cases of interest which we label as $i)$ $\lambda = $ const. 
$ii)$ $ \lambda \to 0$ asymptotically and $iii)$ $ \lambda \to \infty$ asymptotically which
we will discuss in turn.
\subsection{$\lambda =$ const.}
This case implies three the existence of three different possibilities. Firstly we could take $\lambda^2 = 0$, which essentially implies that $\tilde V = 0$
or that the potential is a constant. The second possibility is that $x=0$, but this means that the phantom field is constant and doesn't permit dynamical solutions.
The third and final possibility is that $\Gamma = 1$ which uniquely fixes the form of the potential to be of the form 
$V(\tilde \phi) = A \exp (\alpha \tilde \phi)$, where $A, \alpha$ are positive or negative constants of the appropriate dimension.
A nice illustration of the expected oscillatory behavior is shown in figure 3, where we use the potential 
$V \sim e^{-\phi}$ to illustrate the general behaviour. 
Note that the Hubble parameter initially undergoes a bounce, before settling into an oscillatory phase \footnote{This was also discussed in \cite{oscillatinghubble}.}. The amplitude of the oscillations varies rapidly between $-1/2 \le \tilde H \le 1/2$. Similarly the energy density settles into an oscillatory phase, after initially increasing towards its maximal value. It is important to note that the Hubble parameter is approximately constant or slowly decreasing for much of the initial evolution, therefore there will be virtually no corresponding change in the amount of gravitational wave production. Eventually the Hubble parameter reaches zero, at which point is moves into its oscillatory
period (with a fixed amplitude), but oscillating about the zero point. This will therefore lead to an oscillatory signal for the gravitational
waves produced, with an ever decreasing period. This oscillation is expected to be eternal \cite{burin}.
Physically the bounce is due to the way the energy density is distributed into kinetic and potential energies. The potential energy
of the field is always positive, and initially dominates the negative kinetic energy of the rolling field, which is why the energy
density is an initially increasing function. However eventually the field starts to roll rapidly and therefore the kinetic energy
contribution starts to become dominant. Oscillation in the velocity thus makes the total energy density oscillatory, which in turn
forces the Hubble parameter to oscillate about the bounce point.

\begin{figure}[ht]
\centering
\includegraphics[width=13cm, height=9cm]{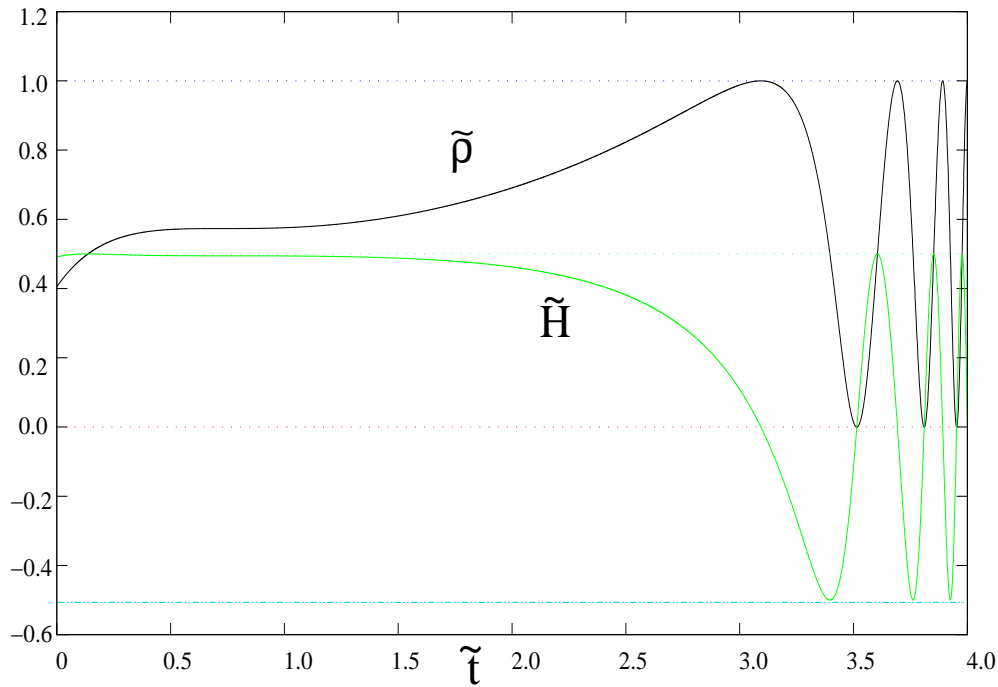}
\caption{Oscillation of the Hubble parameter and energy density when we take $\lambda =$ const, using a potential given by $V = e^{-\phi}$.}
\end{figure}

Let us initially assume that $x_c = 0$. In fact the only fixed point solution can be see to be $x_c=0, y_c=0$ - which is essentially the most
trivial case because the energy density is zero. We can actually solve the equations of motion in this instance. We see that $y$ is constant, and that the boundary term
for the velocity must satisfy the equation $x_0 = \lambda y^2 (t-t_0)$.
However the equation of state at this point must be evaluated as a limit $\rm{Lim}_{x,y=0} \Omega_{\phi}$ in order to avoid singularities. In fact this particular limit can be shown to be non-existent when one applies the tests for multivariable functions, therefore the equation of state is undefined in this instance.

Let us now consider solutions with $y_c=0$, which implies that the fixed point is again at $(0,0)$ if we assume that the phantom field doesn't have complex
velocity. The more interesting case is to set $\lambda=0$, which implies that the following critical points exist $(0,y), (\pm \sqrt{2}y,y)$ and $(\pm \sqrt{2(y^2-1)},y)$,
with $y$ taken to be arbitrary. The point $(0,y)$ admits a constant equation of state - namely $\Omega_\phi = -1$ consistent
with a dark energy dominated phase rather than a phantom phase. The point $(\pm \sqrt{2}y,y)$ leads to a divergent equation of state with $\Omega_\phi \to -\infty$
for all non-zero positive $y$. The final fixed point actually leads to the solution $\Omega_{\phi} = 1 - 2y^2$ and therefore to a whole range of phantom like
behavior provided that $y^2 > 1$. 
Solutions to the equation of motion can be obtained numerically, however they show that $x \to 0$ as a function of time, and that $\Omega_{\phi} \to -1$ from 
below.

Let us now examine the stability of the solution by considering perturbations around the critical points of the dynamical equations,
where we assume the perturbation is small
\begin{equation}
x = x_c + \delta x, \hspace{1cm} y = y_c + \delta y.
\end{equation}
We wish to determine the perturbed equations of motion to leading order. The only non-trivial expression we need to determine is the perturbed
Friedmann equation. If we denote $\tilde H_c^2 = \tilde H^2(x_c, y_c)$, i.e the Friedmann equation evaluated at the fixed points - then we can write
the perturbed solution as follows
\begin{equation}
\tilde{H} \sim \pm \tilde{H}_c \left(1+\frac{(x \delta x - 2y \delta y)(2 y_c^2-1-x_c^2)}{2 \tilde{H}_c^2} + \ldots \right)
\end{equation}
We now insert this into the original equations of motion and expand to linear order in the perturbations. We can write the resultant differential
equations for the perturbed solutions as a matrix equation where we define a perturbation vector $\delta {\bf Z} = (\delta x, \delta y)$, and a perturbation matrix $\textbf{M}$ satisfying
\begin{equation}
\frac{d}{d\tilde t}\left(\delta {\bf Z} \right) = \mathbf{M} \left(\delta {\bf Z} \right),
\end{equation}
whose eigenvalues determine the stability of the solution in question.
A brief calculation yields the following solution for the eigenvalues. Let us first define the function $F(x,y)$ through the 
relation
\begin{eqnarray}
F(x_c,y_c) &=& 9 \tilde H_c^4 -12 \tilde H_c^3 \lambda y_c -12\tilde H_c \lambda x_c y_c(x_c+2y_c)[2y_c^2-1-x_c^2]+ \\ 
&+& 9[2y_c^2-1-x_c^2]^2x_c^2(x_c+2y_c^2)^2 + \tilde H_c^2 \left\lbrace18x_c(x_c+2y_c[2y_c^2-1-x_c^2]+\lambda^2y_c(x_c+4y_c))\right\rbrace \nonumber
\end{eqnarray}
in terms of which we can write the eigenvalues as follows
\begin{equation}
\beta_{\pm} = -3 \tilde H_c^2 -3 x_c(x_c-2y_c)[2y_c^2-1-x_c^2]-2\tilde H_c \lambda y_c \pm F^{1/2}(x_c,y_c)
\end{equation}
The system can be regarded as being stable (at least to linear order in the perturbations) when both eigenvalues are negative definite
In the first instance where we have $(x_c, y_c) = (0,0)$ the eigenvalues are in fact undefined, since there is no possible path through
phase space that leads to a convergent solution. This should have been anticipated from the form of $\Omega_{\phi}$, which 
cannot be defined in this limit.

The non-trivial case where we have $\lambda = 0$ leads to different solutions. For the point $(0,y_c)$ we find that the 
eigenvalues are $\beta = 0, -6 \tilde H_c^2$ implying that this point is a stable node
provided we take the positive sign for the Hubble parameter. The remaining fixed points correspond to $(\pm \sqrt{2}y_c,y_c)$ 
and $(\pm \sqrt{2(y_c^2-1)},y_c)$ and yield zero eigenvalues. This means that they are both saddle points in phase space.
We summarise the results in this section in the following table:
\begin{center}
\begin{tabular}{|c|c|c|c|}
\hline
Validity of $\lambda$ & Fixed points & Eigenvalues & $\Omega_{\phi}$ \\
\hline
$\lambda =$ arbitrary & $(0,0)$ & undefined & undefined \\
$\lambda = 0$ & $(0,y)$ & $\beta_{\pm}= 0 ,-6\tilde H_c^2$ & $\Omega_{\phi} = -1$ \\
$\lambda = 0$ & $(\pm \sqrt{2}y,y)$ & $\beta_{\pm} = 0,0$ & $\Omega_{\phi} \to -\infty$ \\
$\lambda =0$ & $(\pm \sqrt{2(y^2-1)},y)$ & $\beta = 0,0$ & $\Omega_{\phi}= 1-2y^2$ \\
\hline
\end{tabular}
\end{center}
\begin{center}
Table 1. Summary of the fixed points and equation of state for the case of constant $\lambda$.
\end{center}
We also plot the phase space trajectories (Figure 4) for $x$ and $y$ in the specific case where we select $\lambda=0.5$, where we have been
careful to ensure that the trajectories always remain within their regime of validity. This is 
representative of almost all the small, but constant $\lambda$ trajectories.

\begin{figure}[ht]
\includegraphics[width=14cm, height=10cm]{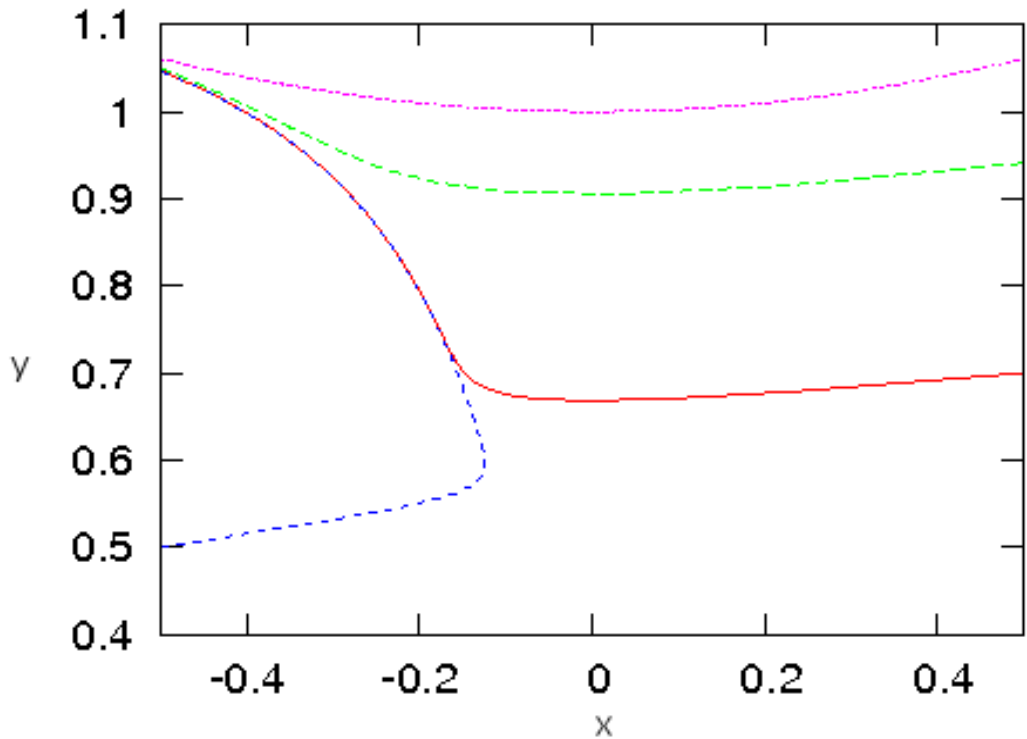}
\caption{Phase space trajectories for values of $(x,y)$ when $\lambda=0.5$, each with differing initial conditions. The upper curve 
corresponds to $y=\sqrt{1+x^2/2}$, which would be the fixed point solution in the limit that $\lambda \to 0$.}
\end{figure}

What is clearly visible from this plot is the trajectories appear to converge at the point $y\sim 1.05$. Note that the top line in
the plot corresponds to the curve $y = \sqrt{1+x^2/2}$. The other curves tend to approach this one at late times, despite having vastly different initial conditions. Of course this will only be an approximate relation since $\lambda$ is small, but non-zero. If we insist
on setting this to zero, then we see that the trajectory is simply a straight line which we have opted not to plot. A similar 
thing happens when we take the converse limit and assume $\lambda >> 1$. The resulting trajectories simply collapse onto a straight line as
you increase $\lambda$ to larger values.

\subsection{$\lambda \to 0$.}
There are a whole class of potentials that satisfy this limit. The simplest being an inverse potential of the form  $V(\phi) = V_0\phi^{-n}$, 
where $V_0$ is an appropriately normalized mass parameter and the phantom is an increasing function of time.
Other examples along these lines correspond to $V \sim V_0 e^{1/\mu \phi}$ and $V\sim V_0 e^{M^2 \phi^2/2}$.
In fact the $\lambda =$ constant solutions discussed in the previous section can be thought of as being instantaneous snapshots of
these solutions. Physically we expect the phase space trajectories, such as those displayed in figure 4, will tend to collapse onto
a single curve as this limit is approached.

The equations of motion in this approximation, provided we ensure that $\Gamma \ne 1$, become
\begin{equation}
\frac{d x}{d \tilde t} \to -3 \tilde H x, \hspace{0.5cm} \frac{dy}{d \tilde t} \to 0, \hspace{0.5cm} \frac{d \lambda}{d \tilde t} \to 0
\end{equation}
Clearly from the first of these equations we see that $\dot{x} < 0$ when $x >0$, and also that $\dot{x} > 0$ when $x <0$ which suggests
that the attractor solution is $x \to 0$ asymptotically. The other expressions again imply that $y$ and $\lambda$ both tend to constants
at late times. How does this affect the expected oscillatory solutions? We see that the frequency will tend to zero at late times regardless of the specific value of $\lambda$ and therefore there will be no oscillation of the Hubble parameter. One can see this intuitively since
the damping factor diverges in this limit (because $x$ eventually settles at the point $x=0$), provided that $y \ne 0$, implying that the solution is overdamped. Note that $\lambda$ is tending to zero, but may remain small but non-zero since it must tend to a constant at late times.

\subsection{$\lambda \to \infty$}
This condition is much harder to investigate because of the equation of motion for $\lambda$. In order for this to be an increasing function of time we must ensure that either $\Gamma < 1$ with $x > 0$, or that $\Gamma >1 $ with $x < 0$. The first of these cases is
not consistent since one can see that this condition implies that $\lambda < -1$, therefore only the second solution is physical. In fact this implies that the potential must follow a power law solution.
From the equations of motion we see that the damping factor increases with time, eventually killing the oscillatory behavior of the solution.

A nice example of potentials satisfying this condition are power law solutions where
$V = M_p^{4-n} \phi^n$. This leads to increasing $\Gamma$ when $n >1$.
Figure 5 illustrates the oscillation of the Hubble parameter for the $\phi^4$ case, where we work explicitly in Planck units. Again we see the bounce solution for the Hubble parameter - and also the initially increasing energy density of the phantom field, before both become oscillatory. The initial increase of both parameters is relatively quick when one compares this to the solution plotted in figure 3, and 
there are significantly more oscillations, each with a reduced period. This is therefore a falsifiable signal of the $\lambda \to \infty$ behaviour. Interestingly the Hubble parameter decreases fairly steadily at early times indicating that the universe continues to expand,
but with a decreasing velocity. This should be contrasted with the plateau phase of the $\lambda=$constant model.

\begin{figure}[ht]\label{{com2}}
\centering
\includegraphics[width=13 cm, height=9cm]{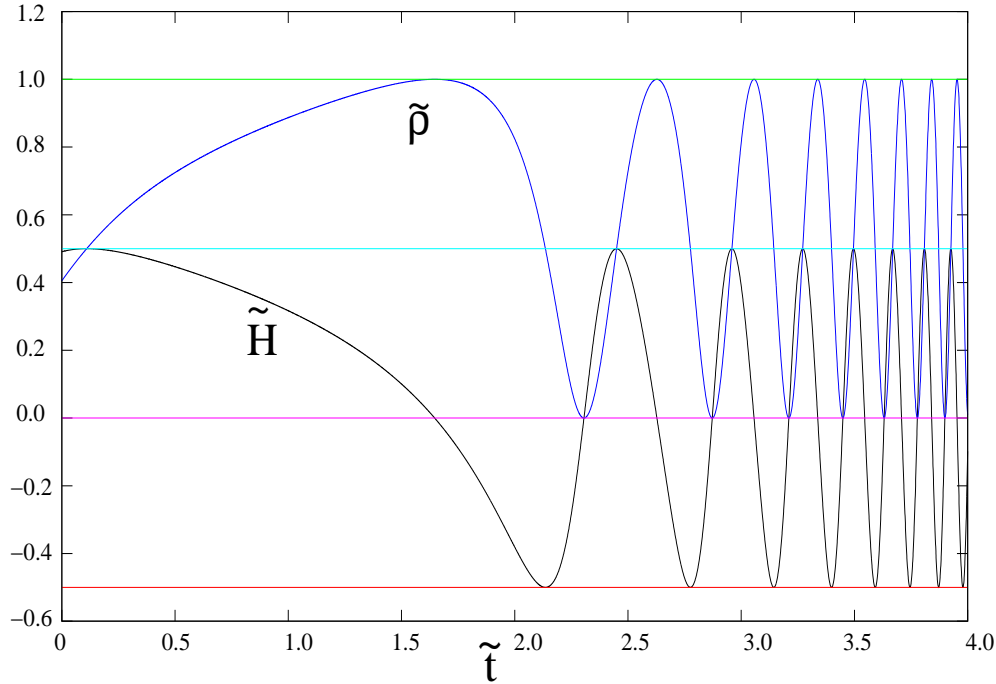}
\caption{Oscillation of the Hubble parameter and energy density assuming a simple $\phi^4$ potential. Note that the initial increasing phase is
relatively short lived, giving way to many oscillations with an ever decreasing period of oscillation. This suggests significantly
different behaviour than the $\lambda =$ const. solution illustrated earlier.}
\end{figure}

\section{Phantom Braneworlds.}
Recently \cite{Singh:2006sg} it has emerged that there is a relationship between loop quantum gravity effects and the Randall-Sundrum (RS) \cite{randallsundrum} braneworld. Recall that in the braneworld scenario there is a brane localized in a $5d$ AdS bulk spacetime, where the extra dimension is taken to be large. This was shown to naturally confine Einstein gravity to the world-volume of the brane \cite{braneworlds}. Although motivated by compactifications of the strongly coupled heterotic string \cite{horavawitten}, the braneworld
scenario is not derivable from a fundamental theory. However it does provide us with an important testing ground for model
building, and in the light of the duality with LQC may offer new insight into the problems of loop quantum gravity.
It is not the purpose of this note to review the braneworld scenario, we refer the interested reader to \cite{braneworldreview}.

In fact - the RS Braneworld has the same autonomous equations of motion as in the loop case \cite{braneworlds}, except for the fact that
$\rho_c$ is given by a slightly different expression involving the tension of the brane. Our analysis therefore 
requires us to consider the brane tension as contributing to the critical density.
There is also a change
of sign in the second term in the Friedmann equation so that schematically we have $H^2 \sim \rho (1 + \rho)$, 
and thus the reality bound on $x$ becomes $x \in \pm \sqrt{2}[y;\sqrt{1+y^2}]$.
\begin{equation}
H^2 = \left(-\frac{x^2}{2} + y^2\right)\left(1-\frac{x^2}{2}+y^2\right).
\end{equation}
The expression for the acceleration of the Hubble parameter remains the same except that we must now define the frequency as follows
\begin{equation}
\omega^2 \to 9x^2 \left( 1+ 2y^2 - 2x^2\right) 
\end{equation}
This has important consequences for the behavior of the solution, as we can see that the frequency is only $\mathbb{R}$ in a finite
domain, and that it becomes complex at late times. This means that $H$ will initially try to oscillate, but as time evolves it
reaches the critical point and freezes out at some constant value \footnote{The oscillation continues but in the complex plane}.
This is completely different to the loop behavior and offers us some falsifiable prediction. One can see this since the 
condition for $\omega$ to be complex is when $2x^2 > 1 + 2y^2$, which can lie within the bound coming from the Friedmann
equation, where we must ensure $x^2 \le 2(1+y^2)$. Therefore there is a region of phase space where $H^2$ is real, but the frequency
of the oscillation becomes complex.

We can analyze the attractor solutions in the same way as we did for the loop case (see also \cite{phantombraneworlds}). Since the formalism was already discussed earlier
we will be brief in our description. Let us first consider the case of constant $\lambda$ as before. The equations of motion admit
four critical solutions once again, which are given by $(0,0)$ for arbitrary $\lambda$ and $(0,y_c), (\pm \sqrt{2}y_c, y_c)$, $(\pm \sqrt{2(1+y_c^2)},y_c)$ when we take $\lambda = 0$. Even without analyzing the stability of such solutions we can use the 
effective equation of state to understand the physics. The first solution again leads to an undefined equation, for the same reason
as before. The second solution, with $y$ corresponding to a flat direction, yields $\Omega_{\phi} = -1$ which is pure dark energy, and 
doesn't represent a phantom solution. The final solution yields an equation of state given by $\Omega_{\phi} = 1+2y_c^2$ which is 
always greater than zero. In fact it is only the third of the above solutions leads to phantom type behavior.

To analyze the stability of the solutions we must again calculate the eigenvalues of the perturbation matrix. Once again we define the 
function $F(x_c,y_c)$, which this time becomes
\begin{eqnarray}
F(x_c,y_c) &=& 36\tilde H_c^4 - 48\tilde H_c^3\lambda y_c + 24[1+2y_c^2-x_c^2]\tilde H_c \lambda y_c(x_c+y_c) + \\
&+& 9[1+2y_c^2-x_c^2]^2(x_c+y_c)^2-4\tilde H_c^2 \left\lbrace 9[1+2y_c^2-x_c^2](x_c+y_c)-\lambda^2 y_c(x_c+4y_c)\right\rbrace \nonumber
\end{eqnarray}
and therefore we can write the corresponding eigenvalues as
\begin{equation}
\beta_{\pm} = \frac{3(x_c-y_c)[2y_c^2-x_c^2]+\tilde H_c(-6\tilde H_c -4\lambda y_c \pm F^{1/2}(x_c,y_c)/\tilde H_c)}{4 \tilde H_c}.
\end{equation}
The point $(0,0)$ is the same as in the loop case, therefore we will not comment on it further. The second fixed point is again at
$(0,y_c)$, however this time the analysis is complicated by the form of the Hubble parameter. We see that the eigenvalues become
\begin{equation}
\beta = \frac{3}{4 \tilde H_c} \left(-2\tilde H_c^2 - y_c -2y_c^3 \pm |2\tilde H_c^2 - y_c -2 y_c^3| \right).
\end{equation}
Now if we take $\tilde H_c$ to be positive definite then we find that the two eigenvalues are both negative functions of $y_c$, and
therefore correspond to stable points in the phase space. Conversely if we chose the minus sign for the Hubble parameter, then we 
see that both eigenvalues are positive definite, and this corresponds to a repulsive fixed point.
The remaining two fixed points correspond to the zeros of the Friedmann equation, and much like the loop case lead to eigenvalues that are identically zero. This indicates that they are saddle points in phase space. We again summarise these results in the following table

\begin{center}
\begin{tabular}{|c|c|c|c|}
\hline
Validity of $\lambda$ & Fixed points & Eigenvalues & $\Omega_{\phi}$ \\
\hline
$\lambda =$ arbitrary & $(0,0)$ & undefined & undefined \\
$\lambda = 0$ & $(0,y)$ & $\beta_{\pm}= -2 \tilde H_c^2 -y(1+2y^2) \pm |2\tilde H_c^2 - y(1+2y^2)|$ & $\Omega_{\phi} = -1$ \\
$\lambda = 0$ & $(\pm \sqrt{2}y,y)$ & $\beta_{\pm} = 0,0$ & $\Omega_{\phi} \to -\infty$ \\
$\lambda =0$ & $(\pm \sqrt{2(y^2-1)},y)$ & $\beta = 0,0$ & $\Omega_{\phi}= 1+2y^2$ \\
\hline
\end{tabular}
\end{center}
\begin{center}
Table 2. Summary of the fixed points and equation of state for the case of constant $\lambda$ for the braneworld case.
\end{center}

\section{Conclusion}
In this note we studied the dynamics of a phantom driven universe both in a loop and braneworld background. We found the Hubble parameter $\tilde{H}$ and the dark energy density $\tilde{\rho}$ remain finite in future and hence avoid a type I singularity for the loop case, however in the braneworld case we see that the scale factor can still diverge due to the fact that the frequency of the oscillation becomes complex. We analyzed the fixed points of the equations of motion in both cases and found that there were two non-trivial solutions, and two trivial (in the sense that they correspond to $\tilde H = 0$) solutions.
The first non-trivial solution is undefined in both cases, whilst the second is a stable fixed point with the appropriate choice of sign. The trivial solutions both correspond to flat directions, even though that they are located at different points in the $(x,y)$ phase plane.

Our study of the loop background was for a general phantom potential, however we saw that the equations of motion impose restrictions on the allowed form of such potentials. The most interesting solution corresponds to the $\lambda =$ const, of which the $\lambda = 0$ solution is a subset. Potentials satisfying this condition lead to simple harmonic motion for $\tilde H$ in the far future, and are therefore capable of resolving all future singularities. On the other hand, the asymptotic cases where $\lambda \to 0,~ \infty$ should appear to be oscillatory only for a finite time. Eventually the equations of motion force the damping factor to dominate at late times, and the system will be fixed at some value. Of course this bounce appears to pass through the point $H=0$ which is where we would anticipate quantum gravity effects to become important.

In the braneworld case we can again use $\lambda$ to parameterise the solutions, however the Hubble parameter quickly freezes out to become constant because of the reality constraint on the oscillation frequency - even when $\lambda = 0$. This means that this  the braneworld case is similar to the late time behavior of the asymptotic loop case. In the asymptotic cases $\lambda \to 0, ~\infty $ we see that again the damping factor will dominate at late times, and therefore kill off all oscillatory solutions, exactly as in the loop case.
Therefore we conclude that only when we have constant $\lambda$ can we be
sure of avoiding all future singularities. In the other cases it is likely that
the type II singularity ($a \to a_s, t\to t_s, \tilde \rho \to \tilde \rho_s$) could still occur \cite{typeIsingularity}, although the
sudden singularity \cite{suddensingularities} is regarded as being a far weaker singularity.

The rapid oscillation of $\tilde{H}$ and $\tilde{\rho}$ in the loop case leads to some interesting cosmological phenomenon. This may cause late
time particle production and therefore provides a useful falsifiable prediction compared to the braneworld scenario. There will certainly
be an observable gravitational wave signature due to the oscillatory nature of $H$, however the precise physical predictions are dependent
upon the choice of the potential, and therefore depend heavily upon the physical origin of such a phantom field.
In both cases we have assumed that the evolution of the universe is governed by a transient phantom phase, and that eventually
this phase will come to an end. However this may be problematic, since it was argued in \cite{vikman} that one cannot find
trajectories where the eigenvalues of the metric tensor take the opposite sign \footnote{Many thanks to A. Vikman for pointing this out.}. One way to avoid this is to consider a large number
of phantom fields, in which case it may be possible for the moduli space metric to acquire an opposite sign. A detailed 
analysis of this conjecture would certainly be most welcome.
However it is important to understand the differences between the brane and loop solutions during this
temporary phase of phantom domination.

Assuming that the phantom phase is transient allows us to consider more general corrections to the Friedmann equation, such as those arising in Cardassian models \cite{cardassian} or the DGP model \cite{DGP}. It is important to understand the distinguish between the different physical predictions of these theories in order to understand whether all future singularities can be resolved.

\section{Acknowledgment}
We thank M. Sami, Parampreet Singh and S. Tsujikawa for their comments and
suggestions on this note. 
Tapan Naskar is supported by CSIR, India. JW is supported by a Queen Mary Studentship and by the University of Iceland. 


\end{document}